\documentclass[11pt,a4paper,dvips]{scrartcl}
\usepackage{graphicx}
\usepackage{subfig}
\usepackage{lcd}
\usepackage{lcd_title,ifthen}
\usepackage{mathptmx}
\usepackage{helvet}
\usepackage{amsmath}
\usepackage{amssymb}
\usepackage{color}
\usepackage{units}
\usepackage{url}
\usepackage[ps2pdf,bookmarks=true,breaklinks=true,bookmarksnumbered=true,colorlinks=true,linkcolor=webgreen,citecolor=webred,urlcolor=webblue]{hyperref}
\usepackage{breakurl}	
\definecolor{webred}{rgb}{0.5, 0, 0}
\definecolor{webgreen}{rgb}{0, 0.5, 0}
\definecolor{webblue}{rgb}{0, 0, 0.5}
\makeatletter
\def\url@myurlfontstyle{%
  \@ifundefined{selectfont}{\def\UrlFont{\sf}}{\def\UrlFont{\small\ttfamily}}}
\makeatother
\long\def\symbolfootnote[#1]#2{\begingroup%
\def\thefootnote{\fnsymbol{footnote}}\footnote[#1]{#2}\endgroup} 
\lcdnote{2012-011}
\newlength{\capindent}
\newlength{\capwidth}
\newlength{\figwidth}
\newcommand{\icaption}[2][!*!,!]{\hspace*{\capindent}%
  \begin{minipage}{\capwidth}
    \ifthenelse{\equal{#1}{!*!,!}}%
      {\caption{#2}}%
      {\caption[#1]{#2}}
      \vspace*{3mm}
  \end{minipage}}
\def\tev{\, {\rm TeV}}
\def\gev{\,{\rm GeV}}
\def\gsim{\mathrel{\rlap{\lower4pt\hbox{$\sim$}}
    \raise1pt\hbox{$>$}}}

\begin{document}
\begin{titlepage}
%
\vskip 35mm
%
\mydocversion
%
\title{Hidden-Sector Higgs Bosons \\ at a High-Energy Electron-Positron Collider}
%
\author{Jack H. Collins\affiliated{1} \affiliated{2}, 
             James D. Wells\affiliated{1} \affiliated{3}}
\affiliations{\affiliation[1]{Cavendish Laboratory, University of Cambridge, UK} \\
                  \affiliation[2]{Cornell University, Ithaca, NY, USA} \\
                  \affiliation[3]{CERN, Theory Group, Geneva, Switzerland}}
%
\date{July 16, 2012}
%
\begin{abstract}
\noindent
The possibility of a scalar messenger that can couple the Standard Model (SM) to a hidden sector has been discussed in a variety of contexts in the literature in recent years. We consider the case that a new scalar singlet charged under an exotic spontaneously broken Abelian gauge symmetry mixes weakly with the SM Higgs resulting in two scalar mass states, one of which has heavily suppressed couplings to the SM particles. Previous phenomenological studies have focussed on potential signatures for such a model at the Large Hadron Collider (LHC). However, there are interesting regions of the parameter space in which the heavier Higgs state would be just out of reach for LHC searches if its mass is $\gtrsim$ 1~TeV. We therefore investigate the discovery potential for such a particle at a $3\tev$ electron-positron collider, which is motivated by the recent R\&D developments of the Compact Linear Collider (CLIC). We find that such an experiment could substantially extend our discovery reach for a heavy, weakly coupled Higgs boson, and we discuss three possible search channels.
\end{abstract}
%
%
\end{titlepage}
%
%
\section{Introduction}
In addition to the known particles and interactions of the Standard Model (SM), there are many compelling reasons to consider a `hidden world' of particles. We define a hidden sector generically as any set of particles, in addition to those of the SM, that are not  charged under any of the SM gauge groups. The possibilities for such models are endless, and if they have no testable consequences then such speculation has little scientific merit. However, there do exist a limited number of interactions that could exist between a hidden sector and the SM  that are renormalisable and relevant or marginal, i.e.\ corresponding to terms in the Lagrangian with dimension $\leq$ 4 and which therefore are not suppressed below some energy scale. Thus, it is reasonable to identify and explore such possibilities as potential windows into a hidden sector. One possibility is kinetic mixing of Abelian gauge fields of the hidden sector and the SM~\cite{Kumar:2006gm}, where the resulting phenomenology overlaps with standard $Z'$ physics~\cite{bib:ZPrime}.

A second possibility is a renormalisable interaction between scalar fields in the two sectors. This mixing would be between the SM Higgs boson, which may have already been discovered~\cite{Higgs signal}, and an exotic singlet scalar state that obtains a vacuum expectation value.
This possibility therefore corresponds to an extended Higgs sector, with a new scalar that may be responsible for spontaneous symmetry breaking in the hidden sector. For concreteness, we study a minimal `toy model' of this type -- the hidden abelian Higgs model~\cite{bib:SW05} -- which involves the addition of a new scalar singlet charged under a $U(1)_{hid}$ gauge symmetry. The phenomenological consequences of this model for LHC physics have been well studied, but there are significant regions of the parameter space in which the new Higgs particle would be very difficult to detect at the LHC if its mass is $\gtrsim$ 1~TeV \cite{bib:BCW07}. The purpose of the present work, therefore, is to investigate the possibility of detecting such a particle at the Compact Linear Collider (CLIC), a future electron-positron collider experiment currently under discussion~\cite{bib:CLIC04}. This experiment would have some advantages for detecting such a particle when compared to the LHC, so this possibility is well worth exploring. We present our key findings on the basis of a parton level analysis for the recoil mass spectrum from $e^+e^-H$ production in sections \ref{subsec:ElecRecoil}, the $\nu\bar\nu H\to jjl^+l^-+ME_T$ signal in section \ref{subsec:ZZ},  and finally the $H\to hh\to 4b$ signature in \ref{subsec:hh}.

\section{Hidden Abelian Higgs Model}

We begin with a review of the hidden abelian Higgs model. In this model the hidden sector contains a complex scalar singlet $\Phi_{H}$ which is charged under a $U(1)_{\mathrm{hid}}$ gauge symmetry. The Higgs Lagrangian is
\begin{equation}
\begin{split}
\mathcal{L}_{\mathrm{Higgs}} = &|D_{\mu}\Phi_{SM}|^{2} + |D_{\mu}\Phi_{H}|^{2} +
                      m_{\Phi_{SM}}^{2}|\Phi_{SM}|^{2} + m_{\Phi_{H}}^{2}|\Phi_{H}|^{2}\\                     &-\lambda|\Phi_{SM}|^{4} - \rho|\Phi_{H}|^{4} -
                      \eta|\Phi_{SM}|^{2}|\Phi_{H}|^{2}\\
\end{split}
\end{equation}
with $m_{\Phi_{SM}}^{2}$, $m_{\Phi_{H}}^{2}$, $\lambda$ and $\rho$ all positive (while $\eta$ can take either sign). In this case, the $U(1)_{\mathrm{hid}}$ gauge symmetry is spontaneously broken by $\Phi_{H}$ taking a particular vacuum expectation value (vev) ($\langle\Phi_{H}\rangle = {\xi}/{\sqrt{2}} = {m_{\Phi_{H}}}/\sqrt{{2\lambda}}$). The resultant massless Goldstone mode gets absorbed by the vector boson corresponding to the Abelian gauge symmetry, as in the SM. In the unitary gauge which eliminates the unphysical Goldstone modes, the scalar fields can be written
\begin{equation}
\Phi_{SM} = \frac{1}{\sqrt{2}}
\begin{pmatrix}
  0 \\ v + \phi_{SM}(x)
\end{pmatrix} ,\;
  \Phi_{H} = \frac{1}{\sqrt{2}}(\xi + \phi_{H}(x))
\end{equation}
where v($\backsimeq 246$~GeV) and $\xi$ are the vevs of $\Phi_{SM}$ and $\Phi_{H}$, while $\phi_{SM}(x)$ and $\phi_{H}(x)$ are the massive modes of these fields. However, the mixing term $\eta|\Phi_{SM}|^{2}|\Phi_{H}|^{2}$ means that the physical mass eigenstates $h$, $H$ of the theory are a linear combination of the two
\begin{equation}
\begin{split}
\phi_{SM} = &\cos{\omega}\, h + \sin{\omega}\, H \\
\phi_{H} = -&\sin{\omega}\, h + \cos{\omega}\, H
\end{split}
\end{equation}
where we choose $M_{H} > M_{h}$, with
\begin{gather}
\tan{2\omega} = \frac{\eta v \xi}{\rho \xi^{2} - \lambda v^{2}} \\
M^{2}_{h,H} = (\lambda v^{2} + \rho \xi^{2}) \mp \sqrt{(\lambda v^{2} - \rho \xi^{2})^{2} + \eta^{2}v^{2}\xi^{2}}.
\end{gather}

The Higgs masses and mixing are subject to a number of constraints. Precision electroweak constraints and direct searches play central roles in constraining what the parameter space can be. The model has four free parameters. Starting with two key observables -- $M_{H}$ and $\sin{\omega}$ -- one is  left with two parameters that can be chosen freely. An analysis of the parameter space of the theory \cite{bib:BCW07} has shown that large regions are compatible with the theoretical constraints, including the regions which are explored in this paper. 

For $M_{H} \geq 2 M_{h}$, which we assume throughout this study, the decay $H \to hh$ is possible. The partial width for this decay is given by the tree level formula
\begin{equation}
\Gamma(H \to hh) = \frac{|\mu|^{2}}{8\pi M_{H}}\sqrt{1-\frac{4 M_{h}^{2}}{M_{H}^{2}}},
\end{equation}
where $\mu$ is the coupling associated with the $h^{2}H$ term in the Lagrangian, the expression for which can be found in \cite{bib:BCW07}. $H$ can also decay into the SM particles, predominantly $WW$, $ZZ$ or $tt$, with a width $\Gamma(H \to SM) = \sin^{2}\omega\Gamma_{SM}(H \to SM)$, where $\Gamma_{SM}(H \to SM) \sim M_{H}^{3}$ is the standard tree level result for a SM Higgs of mass $M_{H}$. $H$ could, in principle, also decay to hidden sector particles with a hidden width $\Gamma_{hid}$. This possibility has been discussed in \cite{bib:SW05}, and in section \ref{subsec:ElecRecoil} we briefly touch on a CLIC search channel which could be useful if the hidden width is significant. However, in this paper we assume that hidden decays of $H$ are kinematically forbidden (i.e.\ the mass of the hidden sector products exceeds $M_{H}$) and so only the SM and $H \to hh$ decay channels contribute to the total width. The $\sin^{2}\omega$ suppression of widths is central to this study. A SM Higgs resonance becomes so wide once its mass is in the trans-TeV region that it becomes impossible to find and it can no longer be understood as a particle. This suppression allows us to keep the $H$ resonance reasonably narrow despite its large mass, and therefore to exploit standard search channels which have already been considered for the heavy (but sub-TeV) SM Higgs case.

\section{Heavy Higgs production at CLIC}
CLIC is a proposed linear $e^{+}e^{-}$ collider presently under discussion \cite{bib:CLIC04}. It is proposed to have a nominal center of mass energy up to 3~TeV, and a design luminosity of $10^{35}$~cm$^{-2}$s$^{-1}$  which should result in more than $\sim 500$~fb$^{-1}$ of integrated luminosity per year at peak design. Lepton collision events have the advantage of being `cleaner' than at a hadron collider, however, the added complication of strong beamstrahlung at a TeV-scale lepton collider presents its own set of issues. Beamstahlung is the emission of hard photons from the colliding bunches, a consequence of the bending of particle trajectories in the strong electromagnetic fields which results from the small bunch sizes required in a high luminosity linear collider \cite{bib:YoCh90}. The effects which are most relevant to our study are the resulting $\gamma\gamma \to$ hadron background, and the spread in the center of mass energies of $e^{+}e^{-}$ collisions.

Collisions between beamstrahlung photons result in hadron production focussed mainly in the forward region of the detector, ($\theta \lesssim$ 100~mrad), but also with energy deposition per event $\lesssim$ 50~GeV in the region $\theta >$ 280~mrad relevant to the hadronic calorimeters \cite{bib:CLIC04}. This has the potential to impact jet reconstruction and missing energy measurements, and proper evaluation of this background requires a detector simulation. We have made no attempt to account for this backround in our study, however, we expect its impact to be modest (e.g., see p.553 of \cite{bib:deRo01}). More significant is the effect of beamstrahlung on the luminosity spectrum; electrons loose energy due to photon emission, and the effects on a colliding pair are correlated which makes a parametrisation of the spectrum difficult. We account for this by making use of a luminosity file generated from the program GUINEAPIG \cite{bib:schu99}, which simulates the beam-beam interaction and produces a list of pairs of particle energies which can be used in a Monte Carlo simulation on an event-by-event basis. 



Heavy Higgs production at CLIC proceeds predominantly via vector boson fusion, either by $ZZ$ fusion with final state $e^{+}e^{-}$, or $WW$ fusion with final state $\nu_{e}\bar{\nu_{e}}$. The $e^{+}e^{-} \to Z^{*} \to HZ$ `Higgstrahlung' process, which dominates at a low energy lepton collider like the International Linear Collider (ILC), gets weaker with increasing center of mass energy such that at CLIC for a 1 TeV SM Higgs the Higgstrahlung production cross section is $\sim$ 1~fb compared with 8.4~fb for $ZZ$ fusion and 79~fb for $WW$ fusion \cite{bib:Alta87}. Although $1\tev$ Higgs boson is somewhat meaningless concept in the SM, due to the strong, non-perturbative growth of its couplings at that high mass, the computation of the cross-section at leading order is still useful for our purposes. This is because the state we are interested in -- the lightly mixed-in singlet Higgs boson --has couplings that are small in comparison and perturbative due to the small mixing angle.

In the following sections we investigate three search strategies, focussing on that described in section \ref{subsec:ZZ}. All signal processes are simulated at tree level using the Monte Carlo event generator \textsc{Pythia} 6.4.24 \cite{bib:Pythia}, while background processes are simulated using \textsc{CompHEP} 4.4 \cite{bib:CHEP}. The luminosity spread due to beamstrahlung is included in the \textsc{Pythia} simulations using \textsc{Calypso} \cite{bib:Calypso}. The beamstrahlung tool available in \textsc{CompHEP} is suitable for a low energy lepton collider, but the Yokoya-Chen approxiation which is used breaks down at high energies and so is unsuitable for our study \cite{bib:YoCh90}. The backgrounds plotted in section \ref{subsec:ZZ} do not include the effects of the luminosity spread. As we will discuss later, we consider this as a small systematic error in our subsequent analysis. All signals are simulated in \textsc{Pythia} using \textsc{Calypso} and so include the effects of beamstrahlung (except where explicitly stated otherwise).

\subsection{Electron recoil mass spectrum}
\label{subsec:ElecRecoil}
\begin{figure}[t]
  \includegraphics[width=0.5\textwidth]{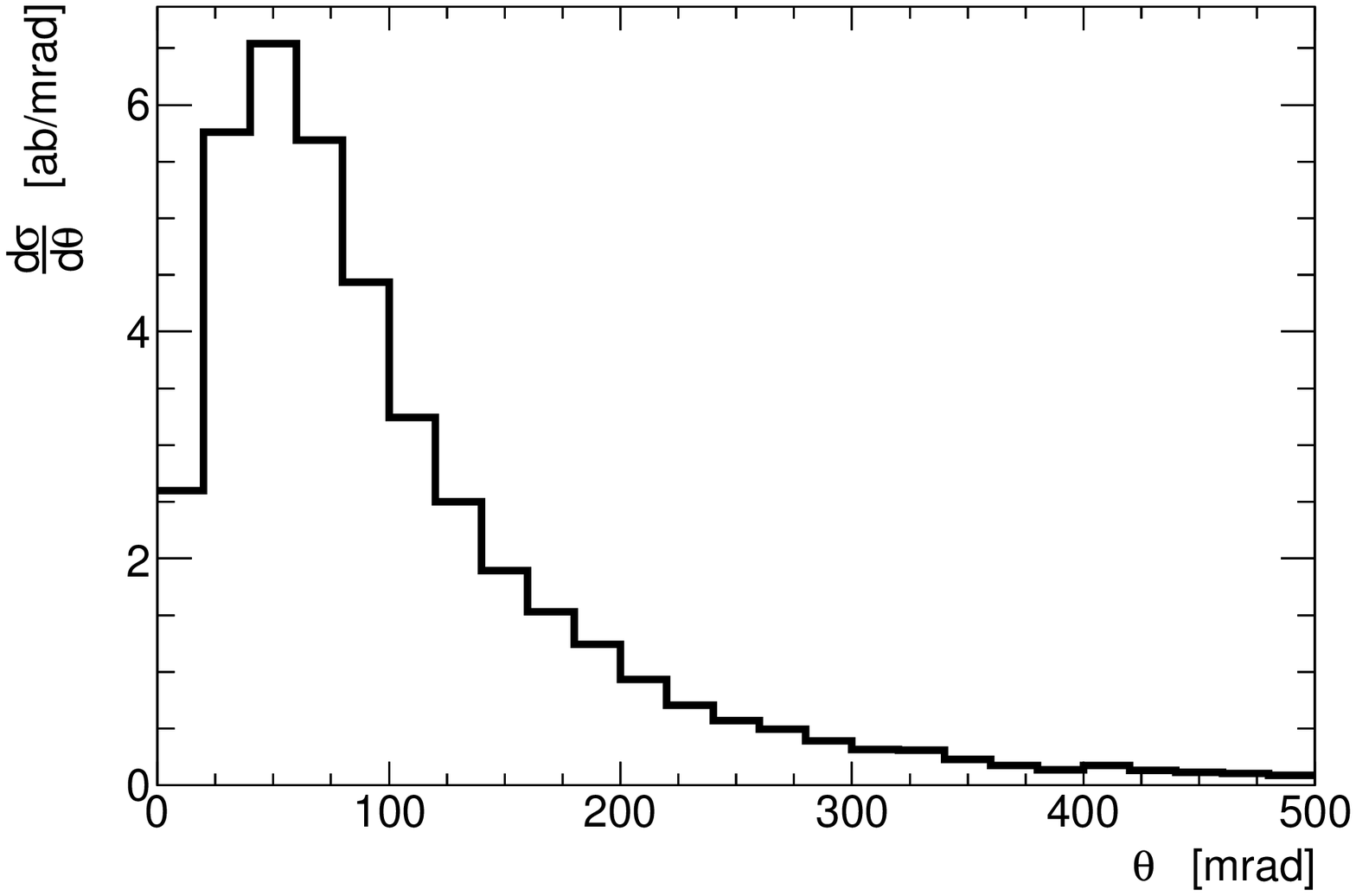}
  \includegraphics[width=0.5\textwidth]{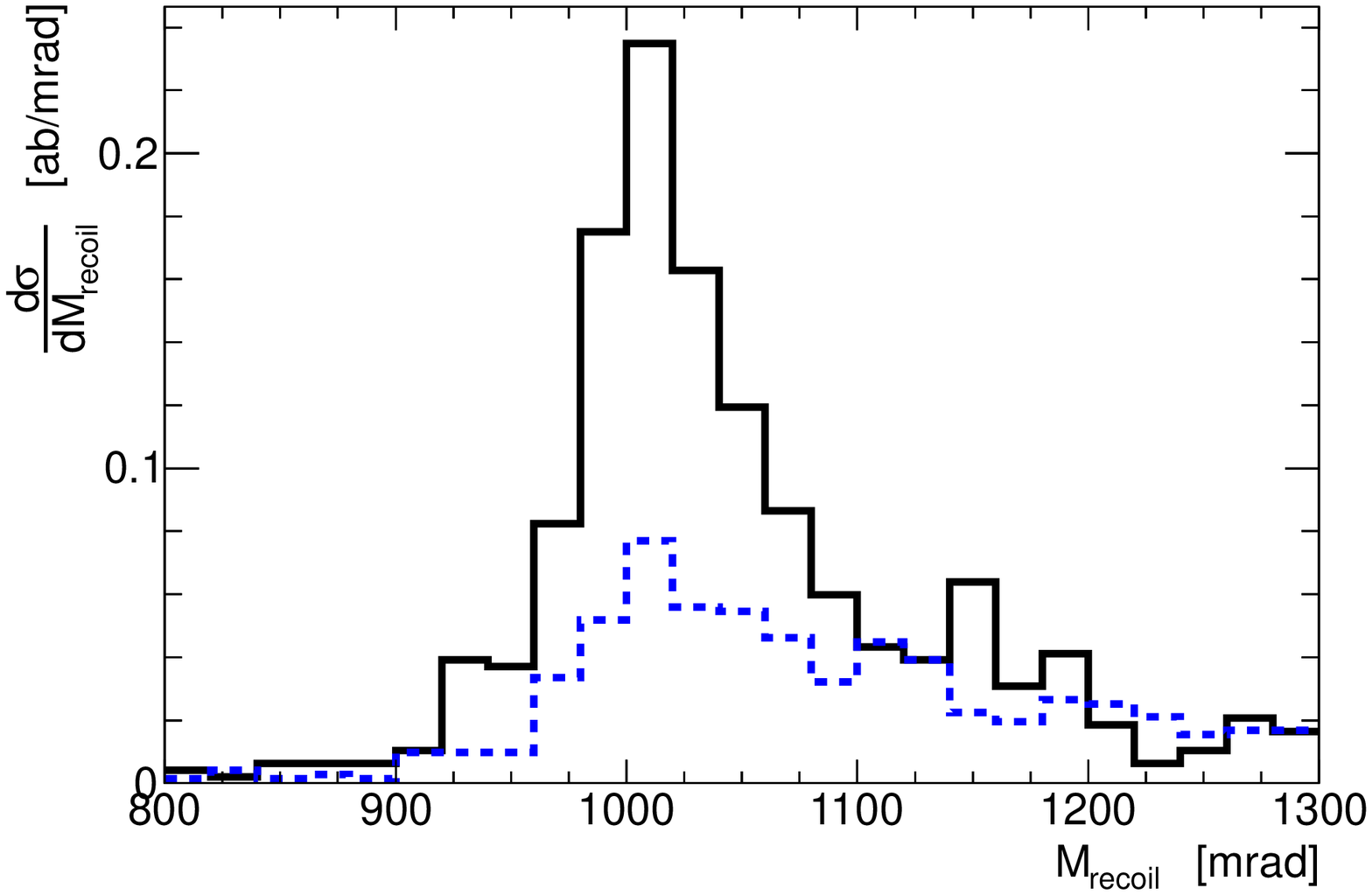}
\caption{Polar angle distribution for recoil electrons (left) and electron recoil mass spectrum (right) for $M_{H}$ = 1 TeV and $\sin\omega$ = 0.3. The dashed blue (solid black) line shows the mass spectrum accounting (not accounting) for the effects of beamstrahlung on the luminosity spectrum. Note that the differential cross section must be multiplied by the bin width (20~GeV in the right plot), to get the cross section per bin.}
\label{fig:recoil}
\end{figure}

One inclusive search strategy is to look for recoiling high energy electrons from the $ZZ$ fusion process, which are peaked in the forward (small polar angle) region. This has the advantage of summing over all possible Higgs decays, even invisible ones, and so provides a rather model-independent search channel \cite{bib:batt01}. It suffers from relying on the smaller $ZZ$ fusion cross section, and from the fact that instrumentation for electron tracking and tagging can only cover angles $\gtrsim$ 100 mrad and therefore can pick up only the tail of the electron distribution which peaks at $\simeq$ 50~mrad (see left panel of figure \ref{fig:recoil}). Events that result in both the outgoing electron and the positron taking an angle $\theta >$ 120~mrad represent only 12\% of the total production cross section.

The electron recoil mass is the invariant mass of the decay products against which the $e^{+}e^{-}$ pair recoil assuming the collision occurs at the nominal center of mass energy, defined as
\begin{equation}
M_{recoil} =  \sqrt{(3000\; \mathrm{GeV} - E_{e^+} - E_{e^-})^{2} - (\mathbf{p}_{e^+} +\mathbf{p}_{e^-})^{2}}.
\end{equation}
A Higgs resonance would potentially show up as a resonant peak in this electron recoil mass spectrum, centered at $M_{H}$. We plot an electron recoil mass spectrum for $M_H$ = 1~TeV, $\sin\omega$ = 0.3 in figure \ref{fig:recoil}, both excluding (solid black line), and including (dashed blue line) the effects of beamstrahlung on the luminosity spectrum. Cuts applied are $\theta_{e^{\pm}} >$ 120~mrad and $E_{e^{\pm}} >$ 700~GeV. Total cross sections for a range of points in the parameter space are presented in table \ref{tab:1}. The width is calculated under the assumption that $H$ decays only into SM particles, though as explained earlier this channel could be used to search for an invisibly decaying Higgs. It can be seen from table~\ref{tab:1} that this channel is promising for a substantial region of the parameter space ($\sin\omega \gtrsim 0.2$ at $M_H \approx$ 1~TeV, $\sin\omega \gtrsim 0.3$ at $M_H \lesssim$ 2 TeV), so long as backgrounds are small, as is expected. It should be noted that for a 600~GeV SM Higgs boson, a consistency-check calculation using this method gives a cross section which is a factor of 4 smaller than given in previous work \cite{bib:batt01}, apparently using the same cuts, and the reason for this discrepancy is unknown.

\begin{table}
\begin{center}
\begin{tabular}{c|c|l|l}
$M_H$ [TeV] & $\sin\omega$ & $\sigma_{0}$ [ab] & $\sigma_{B}$ [ab] \\
\hline
0.6 & 0.5 & 182  & 113  \\
0.6 & 0.1 & 7.31 & 4.80 \\
1   & 0.5 & 73.5 & 38.0 \\
1   & 0.3 & 28.9 & 15.3 \\
1   & 0.1 & 3.24 & 1.61 \\
2   & 0.3 & 7.03 & 3.52 \\
2   & 0.1 & 0.762 & 0.378
\end{tabular}
\caption{Total cross section $\sigma_{0}$ ($\sigma_{B}$) for accepted electron recoil events, ignoring (including) the effects of luminosity spread due to beamstrahlung. The cuts applied are $\theta_{e^{\pm}} >$ 120~mrad and $E_{e^{\pm}} >$ 700~GeV (350~GeV for $M_{H}$ = 2~TeV). These are based on 20000 generated events each, so statistical errors are $\sim$1\%.}
\label{tab:1}
\end{center}
\end{table}

\subsection{$ZZ$ decay channel}
\label{subsec:ZZ}
\begin{figure}[t]
\begin{center}
\includegraphics[width=0.6\textwidth]{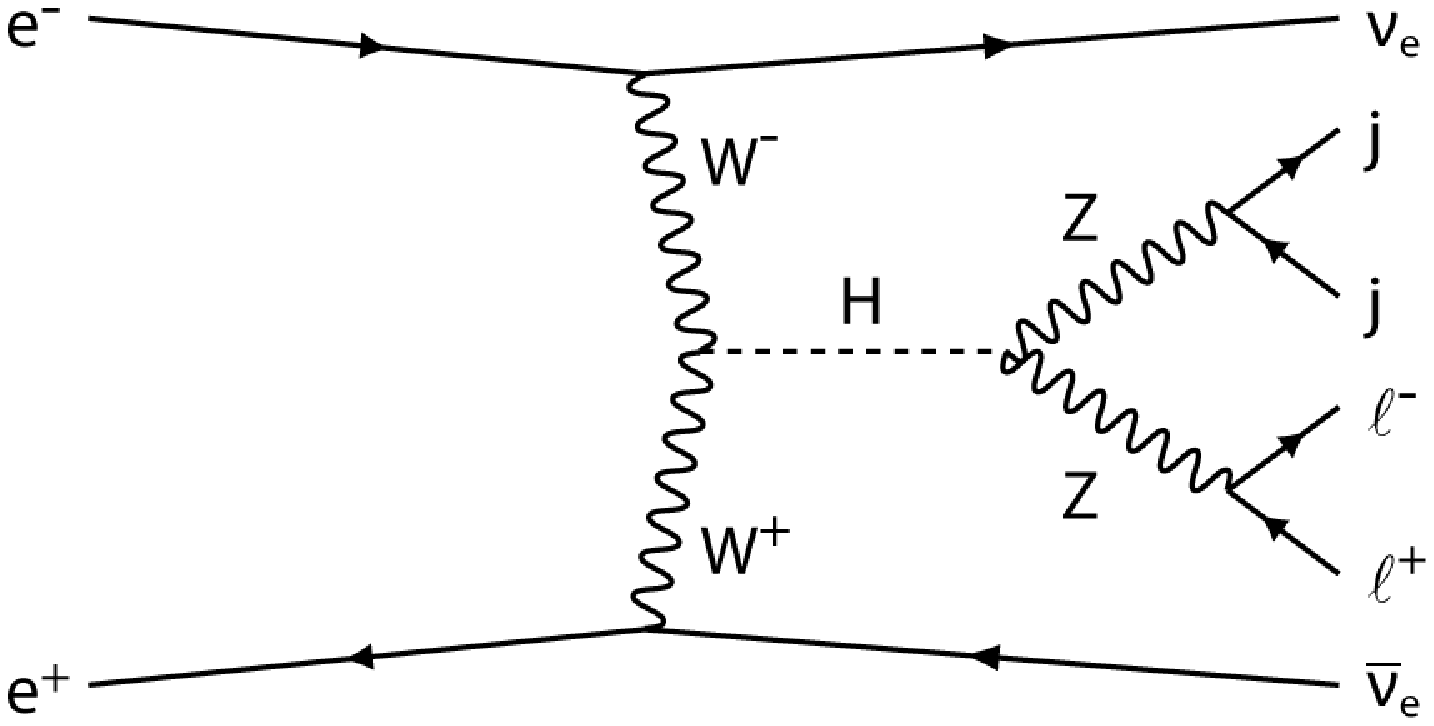}
\end{center}
\caption{$H \to ZZ \to jj\ell\ell$ decay mode.}
\label{fig:feyn1}
\end{figure}

Here we conduct a parton level analysis to investigate the $e^{+}e^{-} \to \nu_{e}\bar{\nu_{e}}(H \to ZZ \to jj\ell^{+}\ell^{-}$) channel (figure \ref{fig:feyn1}), where $j$ represents a quark jet and $\ell$ represents a charged lepton (either $e^{\pm}$ or $\mu^{\pm}$), which benefits from the high $WW$ fusion cross section. We therefore seek events with a jet pair and a lepton pair, each of which has an invariant mass close to that of the $Z$ boson, and then try to identify a resonance peak in the invariant mass spectrum of the $ZZ$ system which would correspond to the Higgs resonance. Missing energy and transverse momentum is provided by the two neutrinos. However, the $ZZ$ fusion process $e^{+}e^{-} \to e^{+}e^{-}(H \to ZZ \to jj\ell^{+}\ell^{-}$) is indistinguishable from $WW$ fusion if the electrons escape down the beam pipe, and this contributes roughly 5\% to the signal. This process is therefore also included in our simulations.

Possible alternatives are 4$\ell$ and 4$j$ signals, however the 4$\ell$ signal suffers from a low $Z \to \ell\ell$ branching ratio which strongly diminishes the signal, while the 4$j$ signal is subject to combinatorial backgrounds and large backgrounds from $e^{+}e^{-} \to e^{+}e^{-}W^{+}W^{-}$ (with the electrons escaping undetected down the beampipe \cite{bib:Barg95}). The $\nu\bar{\nu}jj\ell\ell$ channel therefore represents a good compromise. In our study we conservatively assume a gaussian quark jet energy resolution of 4\% \cite{bib:Thom}, and detector acceptance cuts of $E_{j}, E_{\ell} >$ 20~GeV, $\theta_{\ell} >$ 120~mrad, $\theta_{j} >$ 300~mrad \cite{bib:CLIC04}. We also arbitrarily set the $H \to hh$ branching ratio B($H \to hh$) = 0.04, where BR = $\frac{\Gamma(H \to hh)}{\Gamma(H \to hh)+\sin^{2}\omega\Gamma_{SM}(H \to SM)}$. A very high branching ratio would be likely to result in violation of the perturbative unitarity constraints and result in a strongly interacting Higgs sector, while having the branching ratio be any value less than $\sim 10\%$ would not impact our analysis in a significant way.

The SM $e^{+}e^{-} \to ZZ$ background is very large and potentially drowns out our signal, despite requiring center of mass energies in the low-energy tail of the luminosity distribution in order to compete with the Higgs resonance. However, these $ZZ$ pairs are peaked at $p_{T}(ZZ) =$ 0~GeV, with some spread due to high energy photons being radiated from the initial state $e^{+}e^{-}$ pair which can then escape undetected down the beam pipe. Assuming photon detection capabilities down to 100~mrad polar angle and 10~GeV energy, we find that a transverse momentum cut $p_{T}(\mathrm{vis}) >$ 20~GeV renders this backround negligible compared to other backgrounds at all energies and it is therefore ignored for the rest of this investigation.

We therefore identify the SM processes $e^{+}e^{-} \to \nu\bar{\nu}ZZ$ and $e^{+}e^{-} \to e^{\pm}\nu W^{\mp}Z$ as the principle backgrounds for our signal, where in the latter case the $W$ decays into a pair of quark jets which can be misidentified as a $Z$ due to finite $W$ width and due to the jet energy resolution of the detector, and the outgoing electron escapes undetected down the beam pipe. Diagrams involving colinear photon emission from one of the electrons dominate this $e\nu WZ$ process for low electron angles, and so we use the Weizsaecker-Williams \cite{bib:WWapp} approximation in calculating this background with \textsc{CompHEP}. We find that this calculation differs from that using the complete set of tree level diagrams by no more than a few percent.

\begin{figure}[t]
  \includegraphics[width=0.5\textwidth]{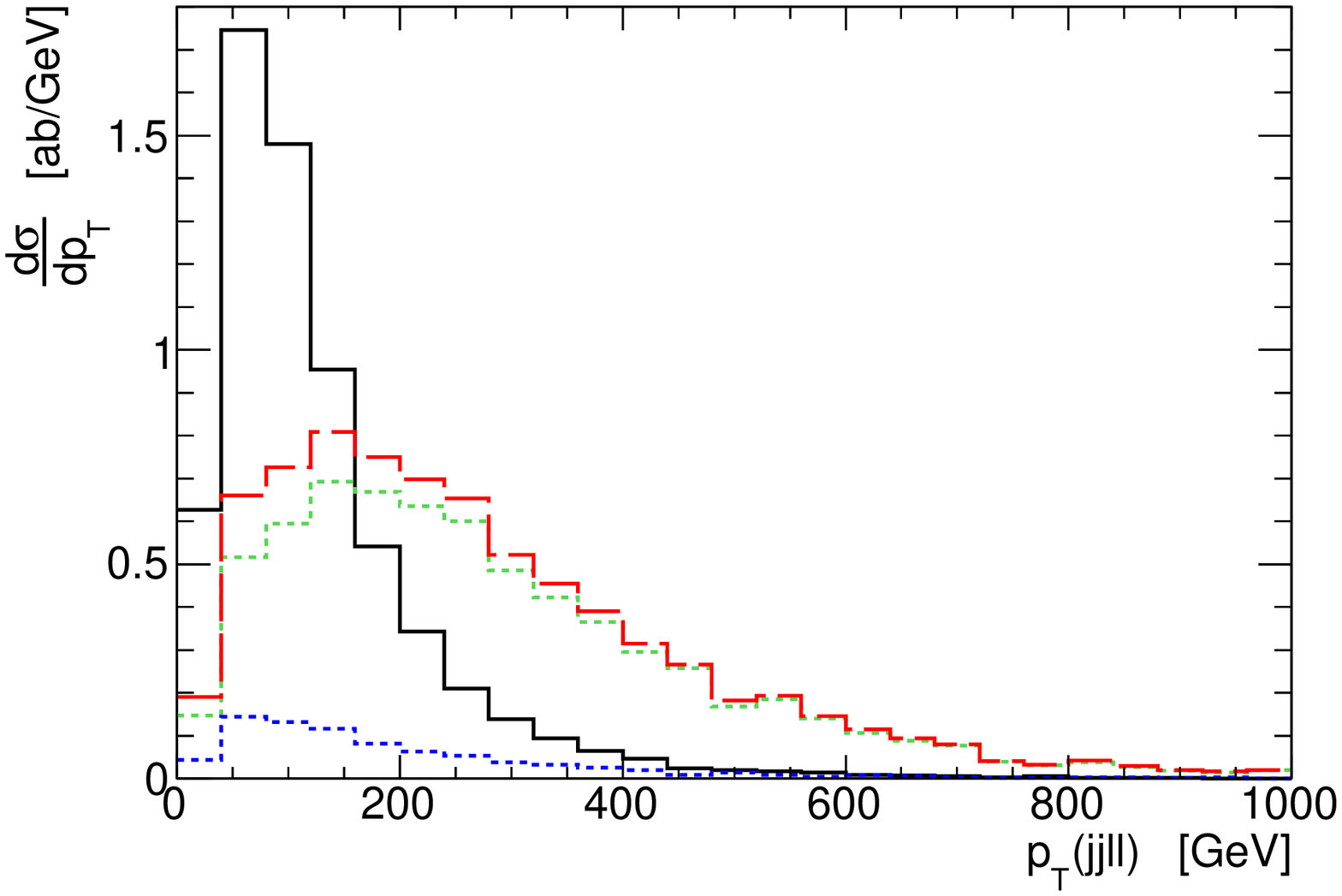}
  \includegraphics[width=0.5\textwidth]{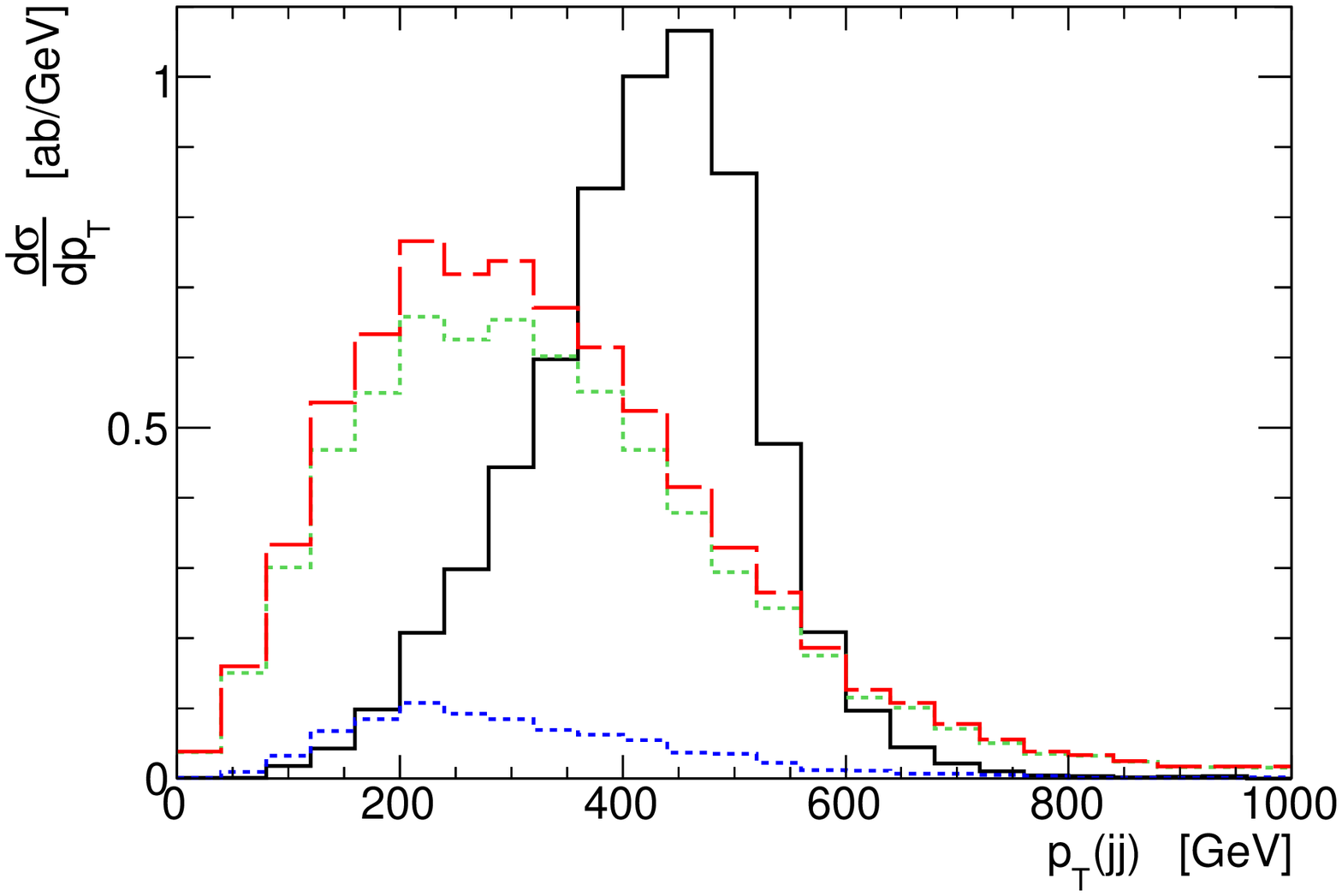}
\caption{$p_{T}$ distributions for the $Z$ decay products of the signal and background for 800~GeV $< M(jj\ell\ell) <$ 1200~GeV. The solid black curves are the signal ($M_{H} =$ 1~TeV, $\sin\omega =$ 0.5), the coloured dotted lines are the two backgrounds (small blue -- $e\nu WZ$, larger green -- $\nu\nu ZZ$), while the red dashed line is their sum.}
\label{fig:ptm1}
\end{figure}

\begin{figure}[t]
  \includegraphics[width=0.5\textwidth]{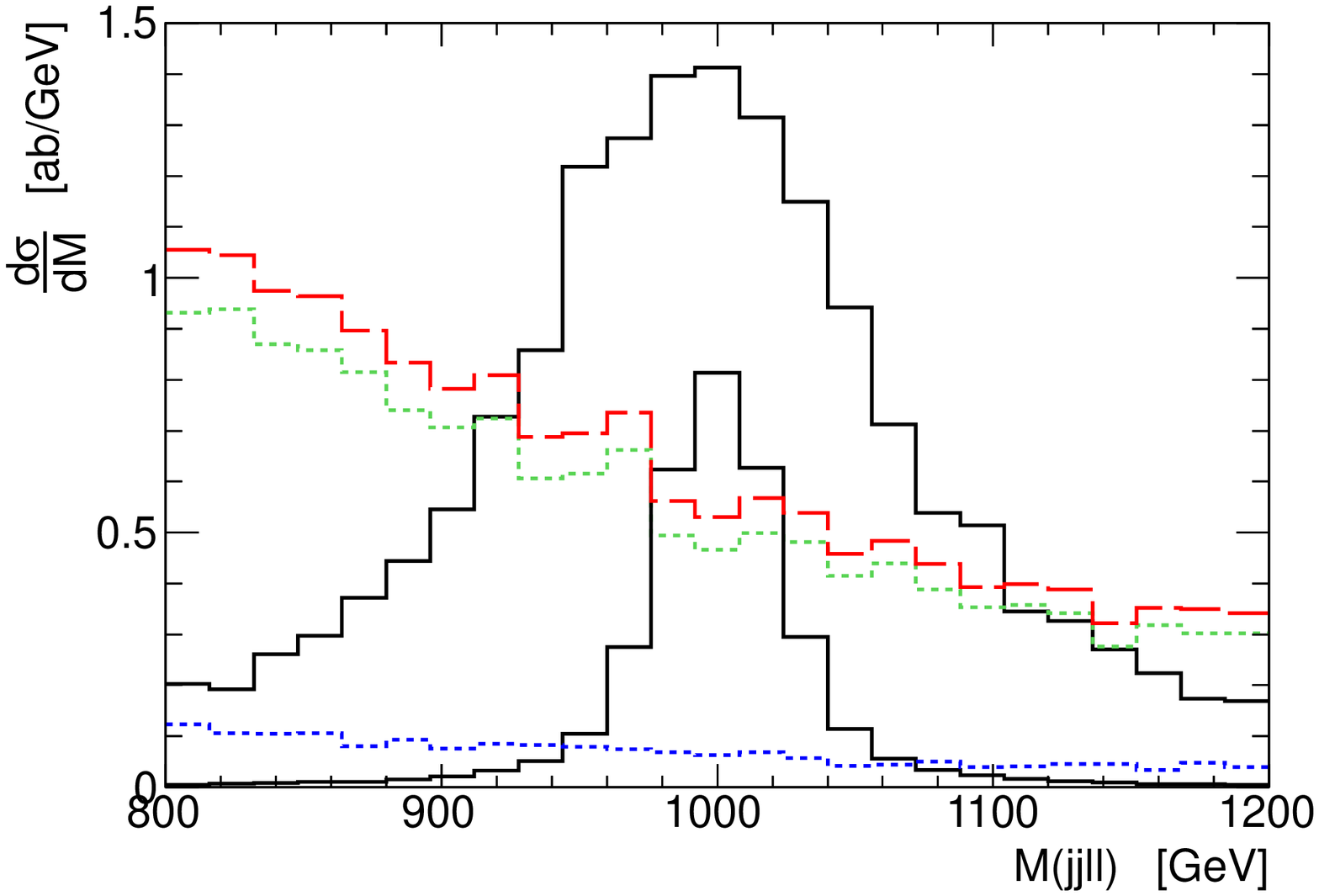}
  \includegraphics[width=0.5\textwidth]{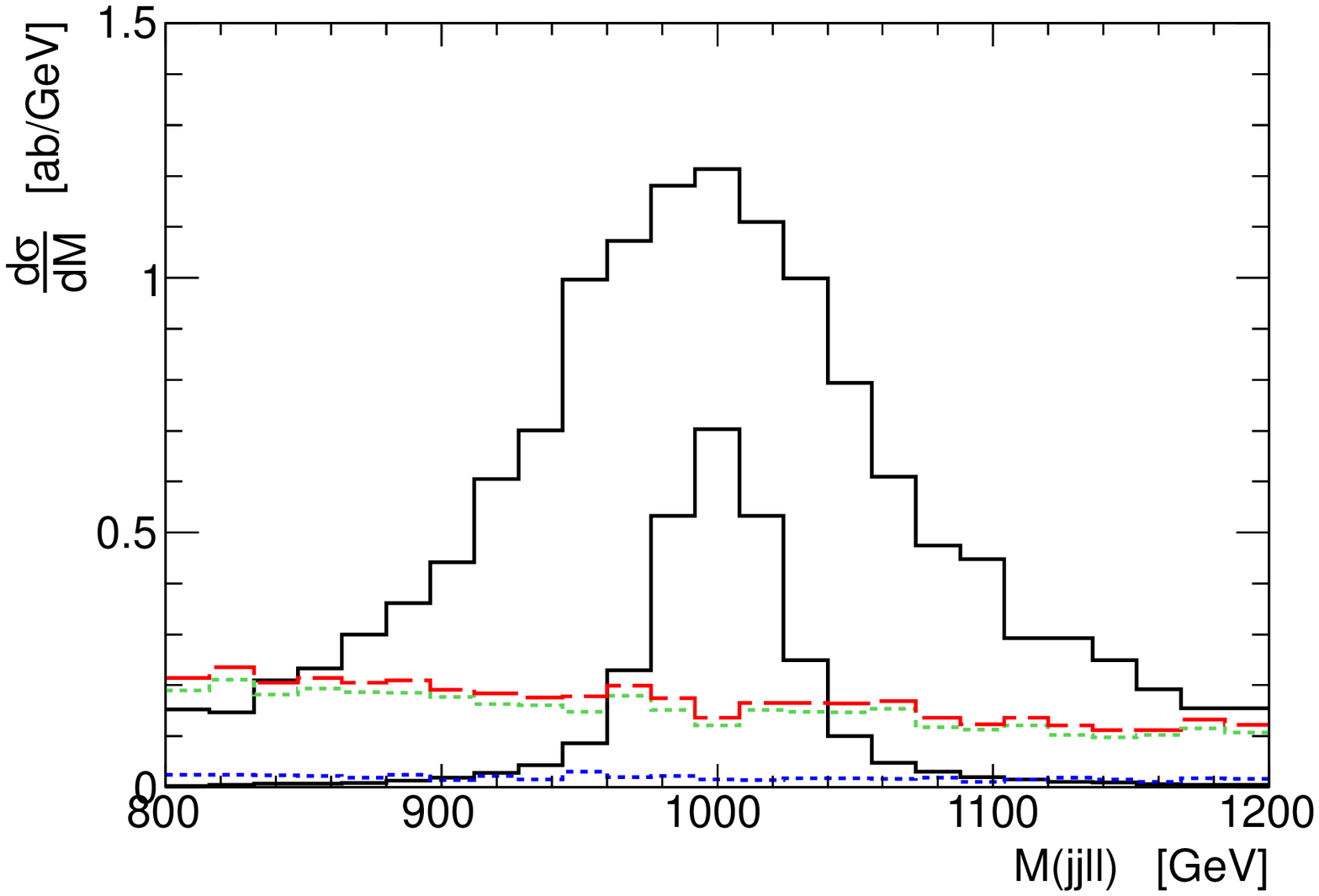}
\caption{Invariant mass distributions of the $M_{H} =$ 1~TeV signal and background before (left) and after (right) applying the $p_{T}$ cuts. The black curves are the signal for two different values of $\sin\omega$, 0.5 for the higher peak and 0.2 for the lower. The blue dotted line represents the $e\nu WZ$ background, green dotted is $\nu\nu ZZ$, and red dashed is their sum.}
\label{fig:mjjllm1}
\end{figure}

These backgrounds can potentially obscure the signal, and to achieve maximum confidence for any discovery we wish to make further cuts in the kinematic distributions in order to suppress the background relative to the signal. The larger of these backgrounds is that from $e\nu WZ$, which is typically a factor of two or three times greater than from the $\nu\nu ZZ$ process (depending on the precise choice of cuts), however good $Z$ identification can reduce it substantially. We therefore choose cuts $\frac{1}{2}(M_{W} + M_{Z}) < M(jj) < (M_{Z} + 15$~GeV), and ($M_{Z} - 10$~GeV) $< M(\ell\ell) <$ ($M_{Z} + 10$~GeV), where the invariant mass $M(12)$ is defined as $M(12)^{2} = (p_{1} + p_{2})^{2}$. The invariant mass distributions $M(jj\ell\ell)$ for a 1~TeV Higgs with $\sin\omega =$ 0.5 and 0.2 are compared with the backgrounds after these cuts in figure \ref{fig:mjjllm1} (left), where it can be seen that the $e\nu WZ$ background has been reduced to a fraction of the $\nu\nu ZZ$ background. Figure \ref{fig:ptm1} shows the transverse momentum distributions (in a bin 800~GeV $< M(jj\ell\ell) <$ 1200~GeV), and it can be seen that the backgrounds typically have higher $p_{T}(jj\ell\ell)$, but smaller $p_{T}(jj)$ than the signal. This is because the vector bosons that are emitted from the incoming electrons in the fusion process tend to be emitted colinear with the beam leading to a low $p_{T}$ resonance, while the final state $Z$ pair are given large momenta due to the high mass of the decaying Higgs. The diagrams that contribute to the background typically involve $t$-channel boson exchange which broadens their $p_{T}(jj\ell\ell)$ spectrum. We can exploit this by applying the cuts $p_{T}(jj\ell\ell) < 300\gev$, $p_{T}(jj), p_{T}(\ell\ell) > 250\gev$ for a 1~TeV Higgs. The effect that this has on the invariant mass spectrum can be seen in figure \ref{fig:mjjllm1} (right).

We now wish to count signal and background events in some $M(jj\ell\ell)$ bin for a range of values of $\sin\omega$. We have $\Gamma \sim \sin^2\omega$, but for small values of $\sin\omega$ the width is dominated by the finite resolution of the detector (corresponding to the jet energy resolution of 4\%). To find an appropriate set of cuts on $M(jj\ell\ell)$ we add these two effects in quadrature, giving:
\begin{gather}
M_{H} - \frac{1}{2}\Delta(\sin\omega) < M(jj\ell\ell) < M_{H} + \frac{1}{2}\Delta(\sin\omega) \\
\Delta(\sin\omega) = \sqrt{a^{2}\frac{\sin^{4}\omega}{\sin^{4}\omega_{0}} + b^{2}}
\end{gather}
where $M_H$ is a selected hypothesized Higgs mass to compare with the data, the parameters $a$ and $b$ are chosen heuristically to match the widths 
of the resonances (see table \ref{tab:cuts}), and $\sin\omega_{0} = 0.3$. The resulting cuts are not necessarily optimized for maximal signal to background ratio, but are nonetheless found to be effective.

\begin{figure}[t]
  \includegraphics[width=0.5\textwidth]{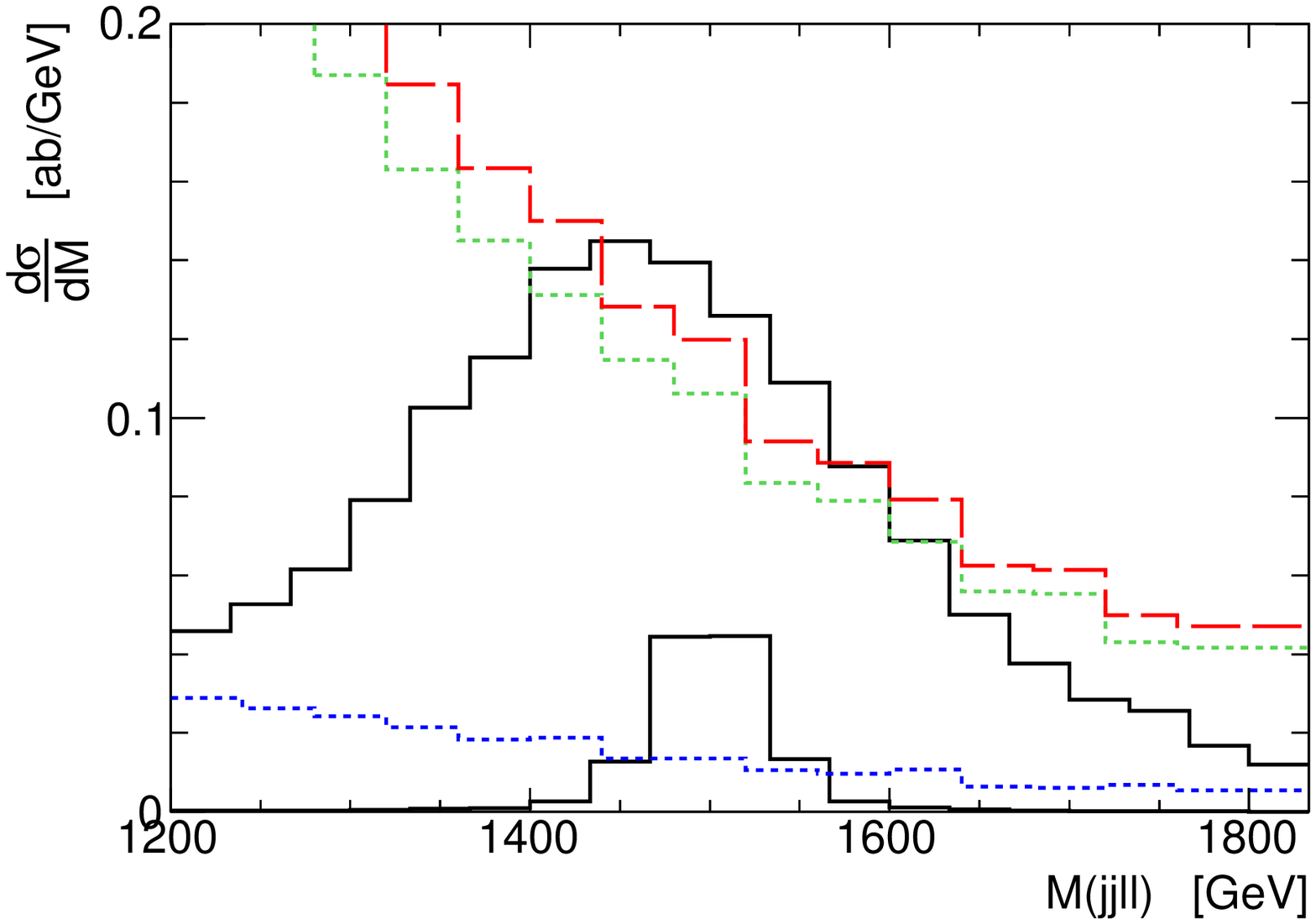}
  \includegraphics[width=0.5\textwidth]{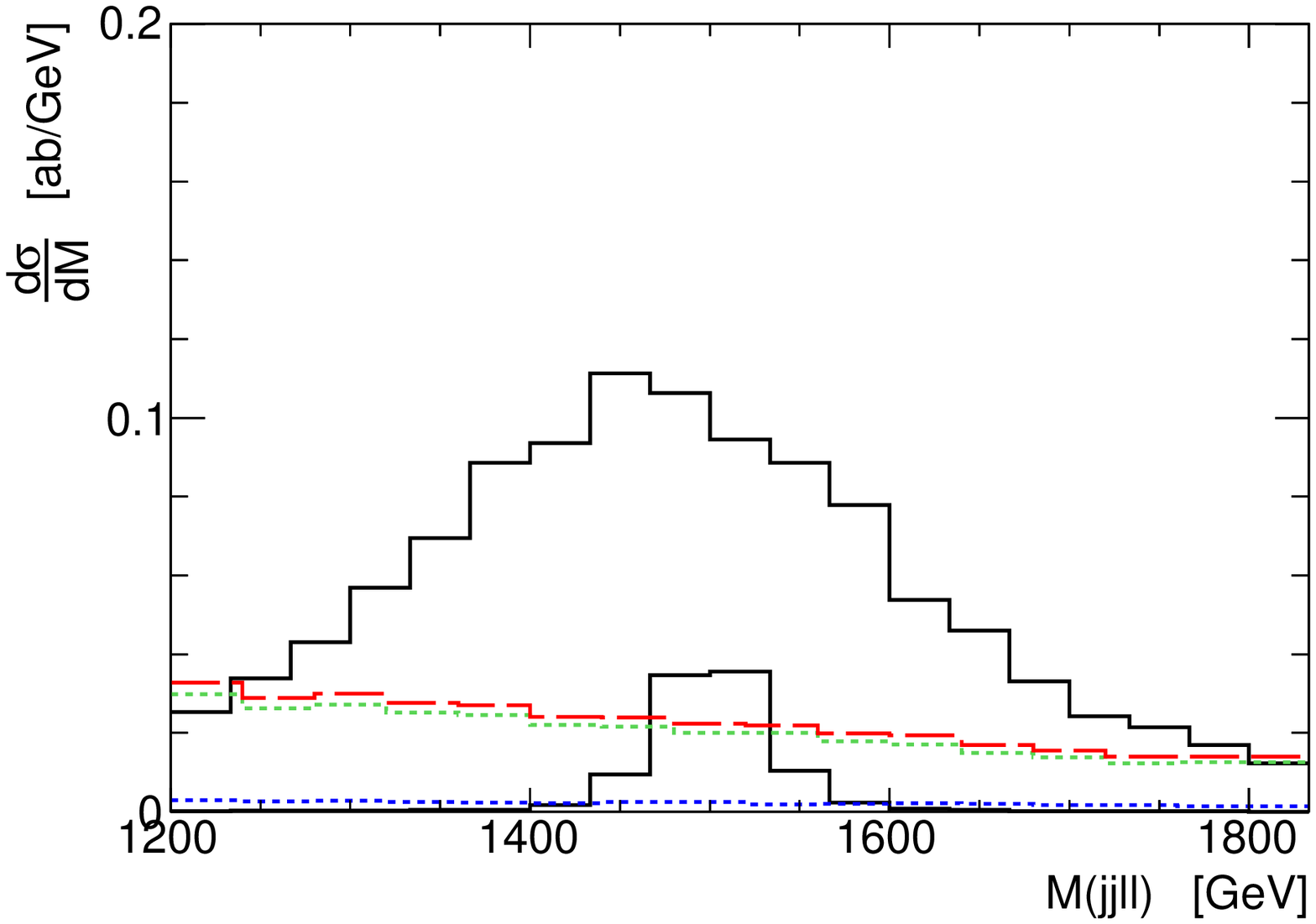}
\caption{Invariant mass distributions of the $M_{H} =$ 1.5~TeV signal and background before (left) and after (right) applying the $p_{T}$ cuts. The signal is represented by a black curve for two different values of $sin\omega$, 0.4 for the higher peak and 0.1 for the lower. The blue dotted line represents the $e\nu WZ$ background, green dotted is $\nu\nu ZZ$, and red dashed is their sum.}
\label{fig:mjjllm15}
\end{figure}

We repeat the preceding analysis to find appropriate sets of cuts for the cases $M_{H}$ = 1.5~TeV and 2~TeV. The cuts which have been chosen are summarized below (equations \ref{equ:cutsstart} to \ref{equ:masscutend}), and the parameters for the different values of $M_{H}$ are given in table \ref{tab:cuts}. The growing value of $p_{T\mathrm{min}}(jj)$ reflects the extra kinetic energy provided by a heavier Higgs resonance, but the declining $p_{T\mathrm{max}}(jj)$ does not reflect any change in the $p_{T}(jj)$ spectrum with $M_{H}$ (it remains roughly constant), but rather reflects a desire for more aggressive cuts as the signal diminishes a little more rapidly than the background as $M_{H}$ is increased (see, e.g.\ figure \ref{fig:mjjllm15} for the $M_{H} =$ 1.5~TeV case).\\

Detector acceptance cuts:
\begin{gather}
\theta_{\ell} > 120 \, \mathrm{mrad}\label{equ:cutsstart} \\
\theta_{j} > 300 \, \mathrm{mrad}\\
E_{\ell}, E_{j} > 20 \, \mathrm{GeV}
\end{gather}

$Z$ reconstruction cuts:
\begin{gather}
\frac{1}{2}(M_{Z} + M_{W}) < M(jj) < M_{Z} + 15 \, \mathrm{GeV} \\
M_{Z} - 10 \, \mathrm{GeV} < M(\ell\ell) < M_{Z} + 10 \, \mathrm{GeV}
\end{gather}

Mass and $p_{T}$ cuts:
\begin{gather}
p_{T}(jj\ell\ell) < p_{T\mathrm{max}}(jj\ell\ell)(M_{H})\label{equ:masscutstart} \\
p_{T}(jj), p_{T}(\ell\ell) > p_{T\mathrm{min}}(jj)(M_{H}) \\
M_{H} - \frac{1}{2}\Delta(\sin\omega, M_{H}) < M(jj\ell\ell) < M_{H} + \frac{1}{2}\Delta(\sin\omega, M_{H}) \\
\Delta(\sin\omega, M_{H}) = \sqrt{a^{2}(M_{H})\biggl(\frac{\sin\omega}{0.3}\biggr)^{4} + b^{2}(M_{H})}
\label{equ:masscutend}
\end{gather}

\begin{table}
\begin{center}
\begin{tabular}{c|l|l|l|l}
$M_H$ [TeV] & $p_{T\mathrm{max}}(jj\ell\ell)$ [$\mathrm{GeV}$]& $p_{T\mathrm{min}}(jj)$ [$\mathrm{GeV}$] & $a^{2}$ [$\mathrm{GeV}^{2}$] & $b^{2}$ [$\mathrm{GeV}^{2}$] \\
\hline
1.0 & 300 & 250 & $1.4\times10^{4}$   & $40^{2}$ \\
1.5 & 250 & 450 & $1.128\times10^{5}$ & $60^{2}$ \\
2.0 & 200 & 650 & $1.456\times10^{5}$ & $120^{2}$
\end{tabular}
\caption{Parameters for the cuts of equations \ref{equ:masscutstart} to \ref{equ:masscutend}.}
\label{tab:cuts}
\end{center}
\end{table}

\begin{figure}[t]
\begin{center}
\includegraphics[width=0.7\textwidth]{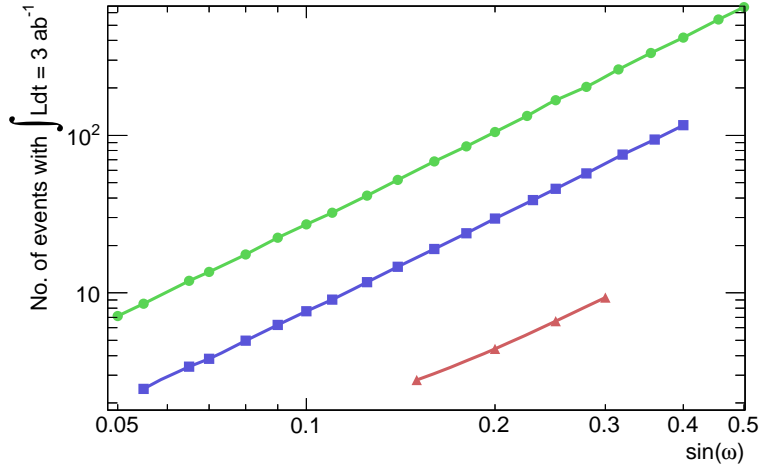}
\end{center}
\caption{Number of signal events from $\bar\nu\nu H\to \bar\nu\nu ZZ\to l^+l^-jj+MET$ that pass the kinematic cuts with $\smallint L dt$ = 3 ab$^{-1}$, with log scales on both the x and y axes. The markers represent the data points. Green circles, blue squares and red triangles correspond to $M_{H}$ = 1.0, 1.5, 2.0 TeV respectively. Each data point is calculated using 10000 signal events, with a statistical error of 1\%.}
\label{fig:numevents}
\end{figure}

\begin{figure}[t]
\begin{center}
\includegraphics[width=0.7\textwidth]{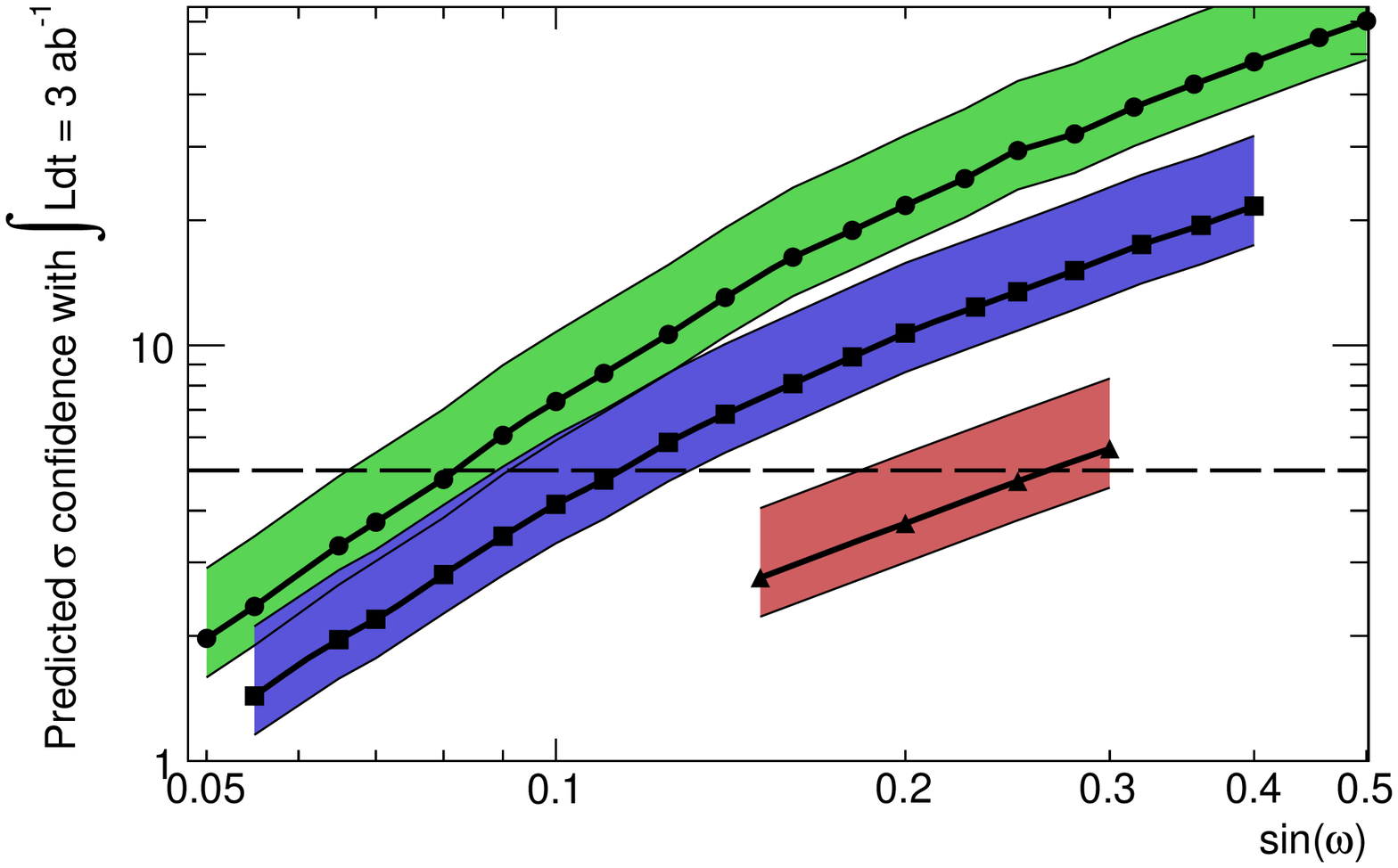}
\end{center}
\caption{Expected confidence level of discovery $\sigma_{\mathrm{conf}}$ with $\smallint L dt$ = 3 ab$^{-1}$ from the signal $\bar\nu\nu H\to \bar\nu\nu ZZ\to l^+l^-jj+MET$. The markers represent the data points. From top to bottom, the curves correspond to $M_{H}$ = 1.0, 1.5, 2.0 TeV. The filled regions represent the systematic error due to the unknown suppression of the background from the beamstrahlung luminosity spread, and are absolute bounds. The upper and lower limits represent beamstrahlung background suppression factors (see text) of 0.3 and 1.0 respectively, and the central line 0.65. Statistical errors are $\simeq$ 3\%.}
\label{fig:sigma}
\end{figure}

The number of signal events passing the aforementioned cuts for $M_{H}$ = 1, 1.5, 2 TeV and for a range of values of $\sin\omega$ between 0.05 and 0.5 are plotted in figure \ref{fig:numevents} with $\smallint L dt$ = 3~ab$^{-1}$, and the corresponding confidence levels are plotted in figure \ref{fig:sigma}. The maximum value of $\sin\omega$ at each mass has been chosen to keep the resonance width within reasonable limits. It has been explained that we have not simulated the effects of beamstrahlung on the background. The cross sections for both background processes diminish with increasing center of mass energy, so the effect of beamstrahlung would be to reduce the size of the background. It can be seen in table \ref{tab:1} that the effect of beamstrahlung on the signal is to multiply it by a factor between $\simeq$ 0.5 and 0.65, depending on the energy scale of the process. The background has a similar energy dependence, so we would expect a roughly similar suppression of the background. Although we cannot be sure of the exact scale of the suppression, we can place an upper bound of 1.0 and a lower bound of 0.3 (since $\simeq$ 30\% of the luminosity is within 1\% of the nominal center of mass energy), and take 0.65 as our expectation. We represent these bounds by the shaded areas in figure \ref{fig:sigma}. For each mass value, the central line represents a 0.65 background suppression factor, the upper confidence bound represents a 0.3 suppression factor and the lower confidence bound represents no suppression (factor 1.0). Although these bounds cover a broad range of confidence levels, we expect the true result not to differ very much from the central expectation value.

\subsection{$hh$ decay channel}
\label{subsec:hh}
\begin{figure}[t]
\begin{center}
\includegraphics[width=0.6\textwidth]{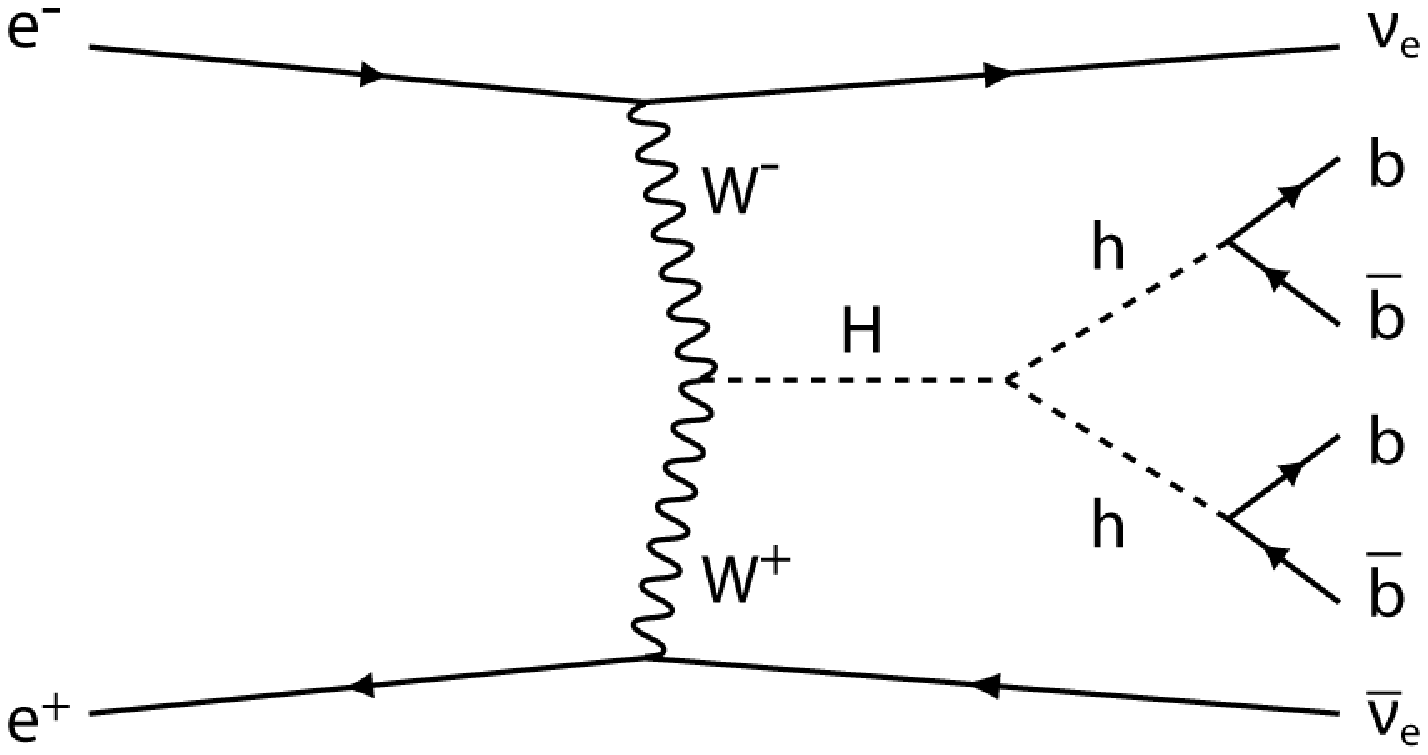}
\end{center}
\caption{$H \to hh \to b\bar{b}b\bar{b}$ decay mode.}
\label{fig:feyn2}
\end{figure}

The decay $H \to hh$ (figure \ref{fig:feyn2}) is allowed if $M_{H} > 2M_{h}$. For the small mixing angles considered here, it can be expected that the light state $h$ will have a mass close to that predicted for the SM, which has bounds 118~GeV $\lesssim M_{h} \lesssim$ 130~GeV from direct searches~\cite{Higgs signal,Kortner:Moriond2012}, with another region opening up for $M_h\gsim 500\gev$, but this region is less favored by precision electroweak analysis. This light Higgs will decay with appreciable branching fraction into bottom quarks $b$. Jets produced from $b$ quarks can leave a highly distinctive signature, and can be identified as such with some probability -- the $b$-tagging efficiency.  The channel $H \to hh \to b\bar{b}b\bar{b}$ therefore has the potential to give a very clear signal, as the backgrounds for $b\bar{b}b\bar{b}$ are likely to be small at an electron-positron collider.

A complete analysis of this would require varying four free observables -- $M_{H}$, $M_{h}$, $\sin\omega$, $B(H \to hh)$, all of which contribute to the size of the signal, and it would be required to test that our points meet the existing theoretical and experimental constraints on the Higgs sector. Instead, we can use the data plotted in figure \ref{fig:numevents} to estimate the number of $b\bar{b}b\bar{b}$ events that we could expect to see, given some value of $B(H \to hh)$ and $B(h \to b\bar{b})$. If we require three or four $b$-tags, the probability of accepting an $hh \to b\bar{b}b\bar{b}$ event is $T(f_b) = 4f_b^{3}(1-f_b) + f_b^4$, where $f_b$ is the $b$-tagging efficiency. A value $f_b = 0.9$ \cite{bib:Thom} gives $T(f_b)= 0.9477$, while a more conservative $f_b = 0.5$ gives $T(f_b) = 0.3125$. The number of $b\bar{b}b\bar{b}$ events, $N_{4b}$, as a function of the number of $jj\ell\ell$ events, $N_{jj\ell\ell}$, is then roughly
\begin{equation}
\begin{split}
N_{4b}& \simeq T(f_b) B(H \to hh) B(h \to b\bar{b})^{2}\biggl(\frac{N_{jj\ell\ell}}{2 B(Z \to jj) B(Z \to \ell\ell) B(H \to ZZ)}\bigg) \\
      & \simeq 13 \, T(f_b) B(H \to hh) N_{jj\ell\ell}
\end{split}
\end{equation}
having taken the values $B(h \to b\bar{b}) = 0.6$ (consistent with $m_h=125\gev$), $B(Z \to jj) = 0.699$, $B(Z \to \ell\ell) = 0.067$, $B(H \to ZZ) = 0.295$. So for instance, with $f_b = 0.9$ and $B(H \to hh) = 0.05$, we get $N_{4b} = 0.62N_{jj\ell\ell}$, while $f_b = 0.5$ would give $N_{4b} = 0.21N_{jj\ell\ell}$. Clearly there are large regions of the parameter space that was explored in section \ref{subsec:ZZ} that could also give a clear $b\bar{b}b\bar{b}$ signal, though this is much more sensitive to the choice of parameters.

\section{Conclusions}
Although current searches for new physics at the TeV-scale are focussed on the experiments at the LHC, it should be remembered that there are possible states that could influence the electroweak symmetry breaking sector yet remain undiscovered there, but which could be discovered by new experiments that we currently have the capability of building. One such possibility is the case where a hidden sector scalar mixes lightly with the SM Higgs, resulting in two physical massive Higgs particles. The light mixing implies a heavily suppressed coupling to the SM for one of these particles, which makes its discovery very challenging.
 
In this paper we have investigated the prospects for finding a trans-TeV Higgs particle at a 3~TeV electron-positron collider and have found that CLIC could substantially extend the reach for such searches compared to the LHC. We have discussed three possible search channels: an electron recoil analysis, which would be particularly useful for an invisibly decaying Higgs, the SM-like decay $H \to ZZ \to jj\ell\ell$ mode which is powerful and generic, and $H \to hh \to b\bar{b}b\bar{b}$ decays which could give a very clear and distinctive signal, shedding more light on the model parameters. Data from a variety of channels could be combined to boost the confidence level for any discovery. It must therefore be stressed that standard heavy higgs search channels that are normally considered for masses $< 1$~TeV could be of importance well into TeV energies, due to the possibility of strong width suppression. Although the case for a multi-TeV electron-positron collider depends on whether new physics is discovered at the LHC and on what form it takes, we believe that the potential for discovering new scalars that couple to a hidden sector, and which might not be found at the LHC, enhances that case. 

\subsection*{Acknowledgements}
We would like to thank M. Battaglia and M. Thomson for helpful discussions.

%
%

\begin{thebibliography}{99}


\bibitem{Kumar:2006gm} 
  J.~Kumar and J.~D.~Wells,
  Phys.\ Rev.\ D {\bf 74}, 115017 (2006)
  [hep-ph/0606183].
  
  \bibitem{bib:ZPrime}
  P.~Langacker,
  Rev.\ Mod.\ Phys.\  {\bf 81}, 1199 (2009)
  [arXiv:0801.1345 [hep-ph]].
  T.~G.~Rizzo,
  hep-ph/0610104.

\bibitem{Higgs signal}
ATLAS Collaboration, ``Observation of an Excess of Events in the Search for the Standard Model Higgs boson with the ATLAS detector at the LHC," http://cdsweb.cern.ch/record/1460439 (9 July 2012); \\
CMS Collaboration, ``Observation of a new boson with a mass near 125 GeV," http://cdsweb.cern.ch/record/1460438?ln=en (9 July 2012).


\bibitem{bib:SW05}
  R.~Schabinger and J.~D.~Wells,
  Phys.\ Rev.\ D {\bf 72}, 093007 (2005)
  [hep-ph/0509209].
  
\bibitem{bib:BCW07}
  M.~Bowen, Y.~Cui and J.~D.~Wells,
  JHEP {\bf 0703}, 036 (2007)
  [hep-ph/0701035].
  C.~Englert, T.~Plehn, D.~Zerwas and P.~M.~Zerwas,
  Phys.\ Lett.\ B {\bf 703}, 298 (2011)
  [arXiv:1106.3097 [hep-ph]].
  C.~Englert et al.,
  Phys.\ Lett.\ B {\bf 707}, 512 (2012)
  [arXiv:1112.3007 [hep-ph]].
  B.~Batell, S.~Gori and L.~-T.~Wang,
  arXiv:1112.5180 [hep-ph].
  
\bibitem{bib:CLIC04}
  L.~Linssen, A.~Miyamoto, M.~Stanitzki and H.~Weerts,
  ``Physics and Detectors at CLIC: CLIC Conceptual Design Report,''
  arXiv:1202.5940 [physics.ins-det].

\bibitem{bib:YoCh90}
K. Yokoya and P. Chen, Frontiers of Particle Beams: Intensity Limitations, eds. M. Month and S. Turner, Lecture Notes in Physics Vol. 400 (Springer-Verlag, Berlin, 1990), pp.414-445.

\bibitem{bib:deRo01}
A. De Roeck, `\textit{WW Scattering at CLIC}', AIP Conf. Proc., 578, 550 (2000).

\bibitem{bib:schu99}
D. Schulte, \textit{Beam-Beam Simulations with GUINEAPIG}, ICAP 1998, CERN-PS-99-014.

\bibitem{bib:Alta87}
G. Altarelli, B. Mele and F. Pitolli, `\textit{Heavy Higgs Production at Future Colliders}', Nucl. Phys. \textbf{B287} (1987) 205.

\bibitem{bib:Pythia}
T. Sjostrand, S. Mrenna and P. Skands, `\textit{Pythia 6.4 Physics and Manual}', JHEP05 (2006) 026 (LU TP 06-13, FERMILAB-PUB-06-052-CD-T) [arXiv:hep-ph/0603175].

\bibitem{bib:CHEP}
E. Boos et al, [CompHEP Collaboration], `\textit{CompHEP 4.4: Automatic computations from Lagrangians to events}', Nucl. Instrum. Meth. \textbf{A534} (2004) 250 [arXiv:hep-ph/0403113];\\
A.S. Belyaev et al, `\textit{CompHEP - PYTHIA interface: integrated package for the collision events generation based on exact matrix elements}', Proc. of ACAT'2000, Fermilab, 16-20 October 2000, p.211 [arXiv:hep-ph/0101232].

\bibitem{bib:Calypso}
CALYPSO can be found at \\ http://home.cern.ch/dschulte/physics/calypso/calypso.html

\bibitem{bib:batt01}
M. Battaglia, CERN-CLIC-Note-474 (2001), [arXiv:hep-ph/0103338].

\bibitem{bib:Barg95}
D. Barger et al., Phys. Rev. \textbf{D52}, 3815 (1995).

\bibitem{bib:Thom}
M. Thomson, Private communications, April 2011.

\bibitem{bib:WWapp}
C.F.Weizsaecker, Z. Phys. 88(1934)612; E.J.Williams, Phys. Rev. 45(1934)729; S. Frixione et al., Phys. Lett. B319 (1993) 339.

\bibitem{Kortner:Moriond2012}
S. Kortner, ``Standard Model scalar boson search with the ATLAS detector," and M. Pieri, ``Searches for Standard Model scalar boson at CMS," Rencontres de Moriond Electroweak (7 March 2012), 
http://indico.in2p3.fr/conferenceOtherViews.py?view=standard\&confId=6001

\end{thebibliography}

\end{document}